\documentclass[superscriptaddress,showpacs,preprintnumbers,amsmath,amssymb,twocolumn]{revtex4-1}
\usepackage{graphicx}
\usepackage{verbatim} 
\usepackage{color}
\usepackage[colorlinks]{hyperref}
\usepackage{epstopdf}

\usepackage{dcolumn}
\usepackage{bm}
\usepackage{upgreek}
\usepackage[super]{nth} 
\usepackage{ulem}

\newcommand{\E}{\mathcal{E}}

\begin{document}

\title{Accurate photon echo timing by optical freezing of exciton dephasing and rephasing  in quantum dots}

\author{A.~N.~Kosarev}
 	\affiliation{Experimentelle Physik 2, Technische Universit\"at Dortmund, 44221 Dortmund, Germany}
 	\affiliation{Ioffe Institute, Russian Academy of Sciences, 194021 St. Petersburg, Russia}
\author{H.~Rose}
    \affiliation{Department Physik \& CeOPP, Universit\"at Paderborn, D-33098 Paderborn, Germany}
\author{S.~V.~Poltavtsev}
 	\affiliation{Experimentelle Physik 2, Technische Universit\"at Dortmund, 44221 Dortmund, Germany}
 	\affiliation{Spin Optics Laboratory, St. Petersburg State University, 198504 St. Petersburg, Russia}
\author{M.~Reichelt}
	\affiliation{Department Physik \& CeOPP, Universit\"at Paderborn, D-33098 Paderborn, Germany}

\author{C.~Schneider}
\author{M.~Kamp}
\author{S.~H\"ofling}
	\affiliation{Technische Physik, Universit\"at W\"urzburg, D-97074 W\"urzburg, Germany} 	
 	
\author{M.~Bayer}
 	\affiliation{Experimentelle Physik 2, Technische Universit\"at Dortmund, 44221 Dortmund, Germany}
 	\affiliation{Ioffe Institute, Russian Academy of Sciences, 194021 St. Petersburg, Russia}

\author{T.~Meier}
\affiliation{Department Physik \& CeOPP, Universit\"at Paderborn, D-33098 Paderborn, Germany}

\author{I.~A.~Akimov}
	\affiliation{Experimentelle Physik 2, Technische Universit\"at Dortmund, 44221 Dortmund, Germany}
 	\affiliation{Ioffe Institute, Russian Academy of Sciences, 194021 St. Petersburg, Russia}

\date{\today}

\begin{abstract}
Semiconductor quantum dots are excellent candidates for ultrafast coherent manipulation of qubits by laser pulses on picosecond timescales or even faster. In inhomogeneous ensembles a macroscopic optical polarization decays rapidly due to dephasing, which, however, is reversible in photon echoes carrying complete information about the coherent ensemble dynamics. Control of the echo emission time is mandatory for applications. Here, we propose a novel concept to reach this goal. In a two-pulse photon echo sequence, we apply an additional resonant control pulse with multiple of $2\pi$ area. Depending on its arrival time, the control slows down dephasing or rephasing of the exciton ensemble during its action. We demonstrate for self-assembled (In,Ga)As quantum dots that the photon echo emission time can be retarded or advanced by 5~ps relative to its nominal appearance time without control. This versatile protocol may be used to obtain significantly longer temporal shifts for suitably tailored control pulses.
\end{abstract}


\maketitle

\section{Introduction}

Coherent nonlinear optics involving quantum emitters is an excellent playground for investigating advanced quantum mechanical phenomena~\cite{Scully-Zubairy}.
Ensembles of quantum emitters in solids typically possess significant inhomogeneous broadening of the optical transition frequency. This might be considered as drawback since it leads to rapid dephasing of a macroscopic polarization in the medium. However, the ensemble may be also used to establish unique collective phenomena in the coherent evolution of the system such as superradiance or entanglement~\cite{SF-Raino-18, Entanglement-Simon-17}
Moreover, ultrafast optical control of quantum emitters in solid state systems is possible on picosecond time scales which is attractive for applications in quantum technologies~\cite{MFox-QuantumOptics} where the use of an ensemble is advantageous for establishing efficient coupling with light~\cite{Lvovsky, Moiseev}.

Semiconductor quantum dots (QDs) are outstanding quantum emitters~\cite{Lodahl-QDs, Bayer-QDs}, showing a discrete energy level spectrum due to the three-dimensional carrier confinement. The transition to the elementary optical excitation of an exciton (electron-hole pair) can often be well approximated by a two-level system (TLS), for which the strong Coulomb attraction in QDs leads to a high spectral selectivity for resonant excitation. QD excitons posssess a large oscillator strength, well defined optical selection rules, and long coherence times, which are limited by the radiative decay time of about 1~ns at low temperatures~\cite{Langbein}. This allows one to perform their ultrafast initialization and coherent control with ps- or even sub-ps-optical pulses~\cite{Bonadeo,Akimov}. Rabi oscillations ~\cite{Steel, Zrenner, Ramsay, Suzuki, Kasprzak} and adiabatic rapid passage \cite{Marie,Phillips} were successfully demonstrated on QD excitons. Ramsey fringes and their control by time dependent electric fields were implemented on single QD level, demonstrating the possibility of optical phase manipulation~\cite{Zrenner-PRL,Zrenner-NP}. In contrast, phase control in an ensemble of emitters has remained difficult because of inhomogeneous broadening of the optical transitions due to fluctuations of QD size, composition etc. In particular, it was shown that rapid dephasing has  significant impact even during excitation with ps-pulses, leading to a complex temporal evolution of the coherent optical response~\cite{RabiQD16, RabiQW17}.

 \begin{figure*}[ht]
	\includegraphics[width = 2\columnwidth]{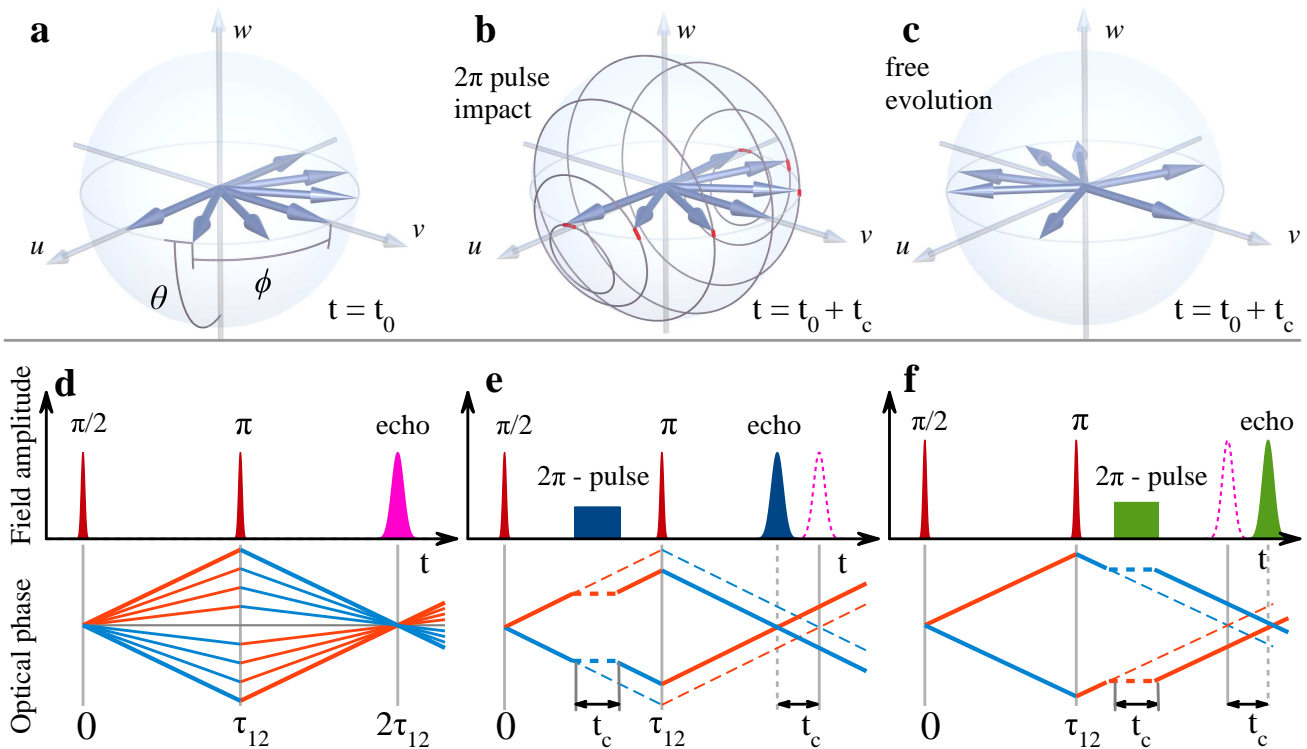}
	\caption{{\bf Photon echo timing} Schematic illustration of coherent evolution of an exciton ensemble on the Bloch sphere. Panel (a) shows partial dephasing of excitons with different detunings at time $t_0$ after excitation with a $\pi/2$-pulse. Panel (b) shows nutation of excitons by the action of control pulse with a pulse area of $2\pi$ (solid gray lines) and weak deviations from pure $2\pi$ rotations due to detunings at time $t_0 + t_C$ with $t_C \approx \Delta^{-1} $, where $\Delta$ corresponds to maximum detuning of the inhomogeneous ensemble (red lines) (c) Distribution of Bloch vectors at the same time as in (b), however, without impact of the control pulse. (d) Phase evolution of excitons in a two-pulse PE sequence from the excitation pulse at $t=0$ to the photon echo at $t=2\tau_{12}$. Positive ($\Delta>0$) and negative ($\Delta<0$) detunings of the exciton resonances are shown by red and blue lines, respectively.  \nth{1} and \nth{2} excitation pulses are shown in red and magenta peak corresponds to the primary two-pulse PE. (e) and (f) Two-pulse PE in presence of the additional $2\pi$ pre-pulse (blue) or post-pulse (green) leading to an advanced and retarded PE peak. The evolution of optical phase in presence of the control pulse is shown by solid lines, while dashed lines indicate the evolution without the control pulse.}
	\label{fig1}
\end{figure*}

The inhomogeneous broadening of optical transitions in a TLS ensemble leads to decay of the macroscopic polarization. However, often the optical coherence on the microscopic level is preserved and the dephasing is reversible. Deterministic delayed emission of light by the QD ensemble can be achieved by the implementation of photon echo (PE) protocols~\cite{Langbein,RabiQD16}. In the simplest case of a primary two-pulse PE, the second excitation pulse is used to invert the phase distribution of the TLS ensemble, which leads to a subsequent rephasing: the appearance of macroscopic polarization and the emission of the PE. Therefore, modification of the TLS phase evolution during dephasing or rephasing must influence the temporal PE profile~\cite{Allen-Eberly}. In nuclear magnetic resonance or electron spin resonance, a large variety of protocols involving complex pulse sequences has been established to shape the dephasing and observe multiple spin echoes on a long time scale~\cite{Slichter}. The most prominent examples are dynamic decoupling (Carr-Purcell and Carr-Purcell-Meiboom-Gill sequences)~\cite{Curr-Purcell, CPMG} and spin locking~\cite{Hartmann-Hahn}. The latter requires elaborate control of the phase in the pulse sequence, which is hard to achieve for optical pulses. Nevertheless, inhibition of dephasing has been also demonstrated for atoms through locking of the macroscopic polarization and the PE shape~\cite{Sleva,Mossberg-1,Mossberg-2}. However, to the best of our knowledge, no optical control of the PE timing has been achieved yet.

In this work, we propose a new approach to this problem and perform a proof of principle demonstration of optical control of PE timing in self-assembled semiconductor QDs, leveraging concepts of multi-wave mixing in TLS~\cite{Rose-Bachelor-work}. First, we show that exciton dephasing and rephasing in a QD ensemble can be efficiently slowed down by resonant excitation with a 2$\pi$ optical control pulse on picosecond timescales. Moreover, we show that the freezing takes place even when the ensemble has lost its macroscopic polarization during the control pulse action. Second, we demonstrate that the application of the control pulse can be used to significantly alter the time of PE emission. Interestingly, this procedure is very robust as there is no need to adjust precisely the optical phase and exact timing of the control pulse. Therefore, our results pave the way to a versatile PE control in ensembles of TLS.

\section{Photon echo timing via dephasing freezing}
\label{sec:Concept}

Here, we describe the main principles underlying the control of PE emission time. Let us consider a TLS ensemble with optically allowed transition between the ground state $|1\rangle_i$ with energy $E_{1,i}$ and the excited state $|2\rangle_i$ with energy $E_{2,i}$. In a semiconductor QD, these states correspond to the crystal ground state and the lowest energy exciton. The energy of this transition is $\hbar\omega_i=E_{2,i}-E_{1,i}$ for the $i$-th QD, which varies due to QD fluctuations. For simplicity, we assume a Gaussian distribution with central frequency $\omega_0$ and inhomogeneous broadening $\Delta_0$. We note that in spite the strong inhomogeneous broadening of the ensemble, i.e., $\Delta_0 \gg 1/T_2$, where $T_2$ is the exciton coherence time in a single QD, the resonance frequency $\omega_0$ is well defined because $\Delta_0 \ll \omega_0$.

In the following, we concentrate on the primary two-pulse PE, resulting from resonant excitation of the ensemble with a sequence of two short optical pulses with areas $\pi/2$ and $\pi$, respectively. The pulse area is defined as
\begin{align}
\Theta=\int_{-\infty}^{+\infty}\Omega_R(t) dt,
\label{pulse-area}
\end{align}
where $\Omega_R(t)=d_{12}\E(t)/\hbar$ is the Rabi frequency with $\E(t)$ being the time-dependent electric field amplitude of an optical pulse and $d_{12}$ is the dipole matrix element of the optical transition. In addition, a control pulse with area $2\pi n$ ($n$ is an integer) is applied in order to freeze the dephasing. The central photon energy $\hbar\omega$ is the same for all pulses and corresponds to resonant excitation, i.e.,  $\omega=\omega_0$. First, let us consider a simplified picture, where the excitation pulses are very short ($\delta$-pulses), while the control pulse has a rectangular shape with duration $t_C$. In this case, the ensemble dephasing during excitation by the first and second pulses can be neglected, simplifying the descriptive analysis. In addition, we neglect primarily any decoherence during PE formation and use the Bloch sphere presentation to describe the coherent dynamics of the TLS ensemble in the rotating frame.

The quantum mechanical state of each QD is given by the coherent superposition of $|1\rangle_i$ and $|2\rangle_i$, and described by the Bloch vector $\mathbf U=(u,v,w)=(\sin\theta \sin\phi, \sin\theta \cos\phi, - \cos\theta)$. The colatitude $\theta$ gives the population difference between the excited and ground state and the longitude $\phi$ determines the phase of the quantum state as shown in Fig~\ref{fig1}(a). Initially, all QDs are in the ground state ($\theta=0$). At time $t=0$, the $\pi/2$-pulse excitation leads to rotation of the Bloch vectors around the $u$-axis to the equator ($\theta=\pi/2$). Then dephasing of the TLS kicks in. Excitons with frequency $\omega_0$ remain at the same point of the Bloch sphere ($\phi=0$), while for other excitons with detuning $\Delta=\omega_i-\omega_0$, the phase evolves linearly with time $\phi = t \Delta $ as shown in Fig.~\ref{fig1}(d). The Bloch vectors of the excited QDs get evenly distributed along the equator (see Fig.~\ref{fig1}(a)). The second $\pi$-pulse, exciting the ensemble at $t=\tau_{12}$, inverts the phase distribution and leads subsequently to refocusing of the TLS at $t=2\tau_{12}$~\cite{Allen-Eberly, Malinovsky}, when all QD Bloch vectors coincide. This is manifested in the recovery of the macroscopic polarization that was induced by the first pulse, and the emission of a PE pulse with duration of about $\Delta_0^{-1}$. In this evolution, two prominent time intervals between excitation and PE emission exist: time range 1 $0<t<\tau_{12}$ for dephasing and time range 2 $\tau_{12}<t<2\tau_{12}$ for rephasing.

\begin{figure*}[ht]
	\includegraphics[width=2\columnwidth]{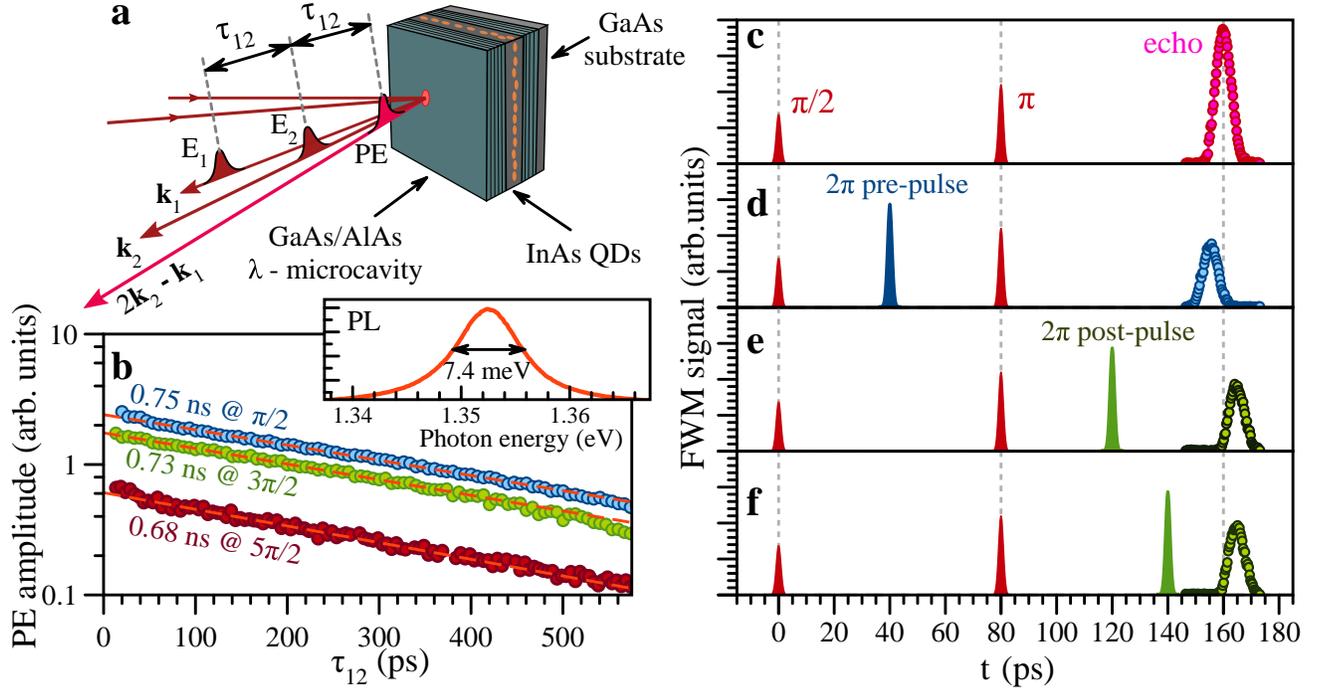}
	\caption{{\bf Photon echo from QDs} (a) Scheme of PE experiment, red arrows indicate the \nth{1} and the \nth{2} pulses and magenta arrow indicates direction of the PE emission. (b) PE decays with increasing delay time $ \tau_{12}$ between pulses \nth{1} and \nth{2} for excitation with different \nth{1} pulse areas $\Theta_1$. Blue, green, and dark red dots correspond to $\Theta_1 \approx \pi/2$,  $3\pi/2$, and $5\pi/2$, respectively. The \nth{2} pulse area is $\Theta_2 \approx \pi$. The inset shows photoluminescence spectrum for non-resonant cw excitation with photon energy  $\hbar\omega_{\rm exc} = 2.33$~eV. (c-f) Temporal profiles of two-pulse PE without control pulse (c), in presence of pre- (d) and post-control pulse (e \& f). \nth{1} and \nth{2} pulses are schematically depicted in red,  pre- and post-control pulses in blue and green. The pulse parameters are $\Theta_1\approx\pi/2$,  $\Theta_2\approx\pi$, $ \tau_{12} = 80$~ps and the control pulse area is $\Theta_C \approx 2\pi$.}
	\label{fig2}
\end{figure*}

Now we apply the control pulse to the dephased system, i.e., when the macroscopic polarization has decayed and all Bloch vectors are evenly distributed along the equator of the sphere. The area of the control pulse is $\Omega_R t_C = 2 \pi n$. In the resonant case ($\Delta=0$), the Bloch vector directed along the $v$-axis is rotated around the $u$-axis, transfering the excitons to the same point on the Bloch sphere after time $t_C$. For detuned excitons, the rotation occurs with the generalized Rabi frequency $\tilde{\Omega}=\sqrt{\Omega_R^2+\Delta^2}$ around the vector $\mathbf\Omega = (\Omega_R,0,\Delta)$, which can be visualized using the equation of motion of the Bloch vector
\begin{align}
\frac{d}{dt} \bm{U} = \bm{\Omega} \times \bm{U} .
\label{bloch-eom}
\end{align}
The analysis shows that there are deviations between the final and starting points of the rotation, as represented by the short red lines in Fig.~\ref{fig1}(b).
For $\Omega_R \gg \Delta$ we obtain $\tilde{\Omega} \approx \Omega_R [1+(\Delta/\Omega_R)^{2}/2]$
(further details are provided in the appendix section ~\ref{Sec:Rect}).
Moreover, since $\mathbf\Omega$ deviates only slightly from the $v$-axis, the increment mostly contributes to $\theta$ rather than to $\phi$. The deviation of the phase after several Rabi rotations from the initial point is significantly smaller compared to free evolution of the system without control pulse after the same time $t_C$ (see Figs.~\ref{fig1}(b) and (c) for comparison). Thus, the control pulse leads to an inhibition of dephasing after its action. Note that there are no strict requirements regarding the duration $t_C$ of the control pulse, the freezing will take place also for $t_C \Delta > 1$. Here, we assumed only that $\Omega_R \gg \Delta$ and $ (\Delta/\Omega_R) t_C \Delta  \ll 1$ (see appendix section \ref{Sec:Rect}).

The impact of the control pulse on the PE formation is manifested in a temporal shift of the PE peak. If the control pulse is applied in time range 1 (control pre-pulse) or 2 (control post-pulse), it effectively slows down the dephasing or rephasing process, respectively. The phase evolutions for these two scenarios are shown in Fig.~\ref{fig1}(e) and (f). A consequence of the partially suppressed dephasing in time range 1 is a shorter rephasing period after the arrival of the second pulse. As a result the PE is advanced by $t_C$ (blue peak in  Fig.~\ref{fig1}(e)). Vice versa, if the rephasing was frozen for the time $t_C$, it will delay the PE appearance after the second pulse by this time (see Fig.~\ref{fig1}(f)).

Interestingly, the effect is independent of the exact timing of the control pulse within the two time ranges, and the optical phase of the control with respect to the first and second pulse is also not important. Thus, this method can be used as convenient and efficient tool to control the timing of two- or even multiple-pulse PEs. In addition, the area of the control pulse can be used to modulate the amplitude of PE. In the following sections, we present experimental data demonstrating that the concept can be easily implemented with self-assembled (In,Ga)As quantum dots. The measurements are analyzed using the optical Bloch equations (OBE) from which good agreement between the experiment is obtained.

\section{Coherent optical response in QDs}

The experiments were performed on a single layer of (In,Ga)As QDs placed at the electric field antinode of a planar $\lambda$-microcavity formed by distributed Bragg reflectors (for details see methods)~\cite{Maier2014}. The quality factor of the microcavity is about 200, as evaluated from the photoluminescence spectrum in the inset of Fig.~\ref{fig2}(b). The photon energy of the optical pulses was tuned into resonance with the photonic mode. The use of the microcavity has several advantages~\cite{RabiQD16, Kasprzak}: First, it allows us to achieve Rabi rotations using moderate excitation intensities. Second, it enhances the intensity of the emitted PEs. Simultaneously, the spectral width of the cavity mode ($\hbar \Delta_B \approx 7.4 $~meV) is significantly larger than the spectral width of the laser pulses ($\hbar\delta\omega \sim 0.6$~meV) and therefore does not influence the coherent dynamics of the optically addressed QD ensemble. On the other hand, we have to consider the QD ensemble as strongly inhomogeneous since the total spectral width of the ensemble $\Delta_0$ is even larger than $\Delta_B$. 

\begin{figure*}[ht]
	\includegraphics[width=1.7\columnwidth]{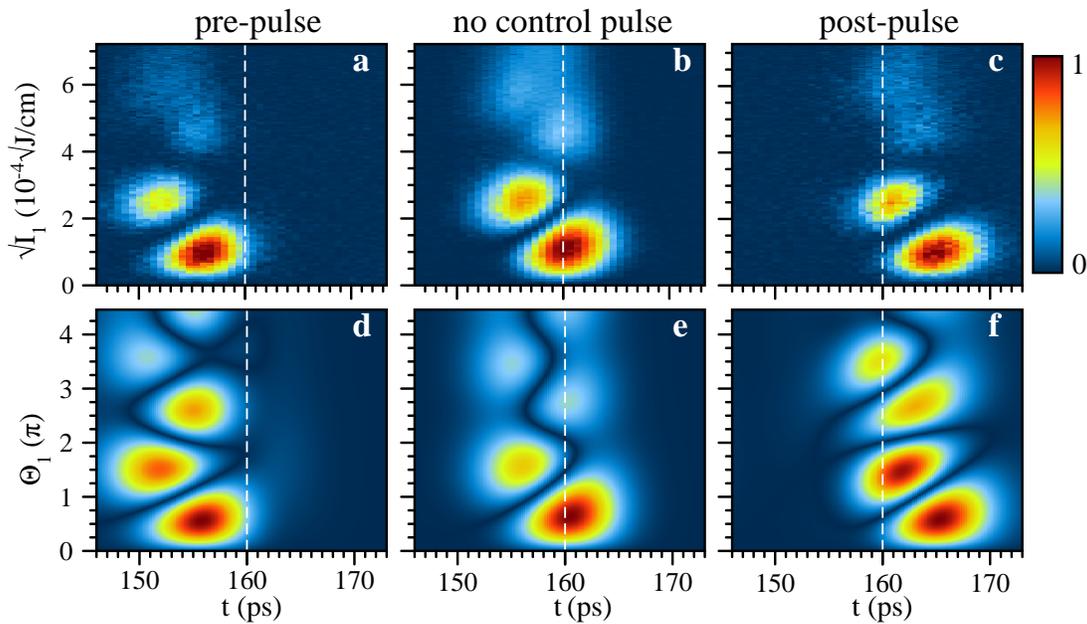}
	\caption{{\bf Impact of control pulse} Two-dimensional plots showing dependence of PE temporal profiles on the amplitude of the \nth{1} pulse $\sqrt{I_1}$ (a,d) in presence of the control pre-pulse, (b,e) without control pulse, and (c,f) in the presence of the control post-pulse. Top row (a-c) and bottom row (d-f) show experiments and simulations, respectively. Data are shown for $ \tau_{12} = 80 $ ps delay, $\Theta_2 \approx \pi$ pulse area of \nth{2} pulse, and control pulse delays of 33~ps relative to the \nth{1} pulse in (a,d) and of 27~ps relative to the \nth{2} pulse in (c,f).}
	\label{fig3}
\end{figure*}

The coherent optical response after the two excitation pulse sequence was measured using transient four-wave-mixing (FWM) in reflection geometry at the temperature of 2~K (see Fig.~\ref{fig2}(a) and Methods). All pulses are emitted from the same mode-locked laser source and have a duration of about 2.5~ps. Heterodyne detection allows us to measure the temporal profile of the electric field amplitude of the FWM signal from the PE~\cite{RabiQD16}. A typical signal is shown in Fig.~\ref{fig2}(c) for a delay time between the \nth{1} and \nth{2} pulses of $\tau_{12}=80$~ps. As expected, the PE peak occurs at the time $t = 2\tau_{12}$. The measured signal is well described by a Gaussian with full-width at half-maximum (FWHM) of about 6.5~ps, which corresponds to a PE duration of 4~ps FWHM (see details in appendix section~\ref{Sec:PE-duration} for evaluation of the PE pulse duration). The PE peak dependence on $\tau_{12}$ is shown in Fig.~\ref{fig2}(b). The PE amplitude decreases with increasing delay time between the \nth{1} and \nth{2} pulse and is well described by $P_{\rm PE} \propto \exp{(-2\tau_{12}/T_2)}$. Interestingly, an increase of the \nth{1} pulse's area $\Theta_1$ from $\pi/2$ to $5\pi/2$ leaves $T_2$ unchanged at $\approx 0.7$~ns. Thus, we conclude that excitation-induced dephasing (EID) is weak in our system, which is essential for robust Rabi oscillations~\cite{RabiQW17}.

The ensemble comprises neutral as well as charged QDs which are occupied with resident electrons. To address only one type of TLS and avoid additional inhomogeneities of the Rabi frequency due to potential differences in the dipole matrix elements for trions (electron-hole pair excitation in a negatively charged QD) and excitons we used a polarization configuration that results in a PE signal from charged QDs only~\cite{Pola19}. This is ensured for a horizontally H-polarized \nth{1} pulse and a V-vertically polarized \nth{2} pulse with the resulting PE signal being H-polarized. 

\section{Impact of control pulse on PE timing}

First, we demonstrate PE timing by a control pulse when the primary echo has maximum amplitude, which is achieved for \nth{1} and \nth{2} pulse areas of $\Theta_1 \lesssim \pi/2$  and  $\Theta_2 \approx \pi$. As will be shown below, the corresponding pulse energy densities are $I_1 = 4~{\rm nJ/cm^2}$  and $I_2 = 23~{\rm nJ/cm^2}$. In this case the PE peak appears at $t=2\tau_{12}$ (see Fig.~\ref{fig2}(c)). The application of a $2\pi$ control pulse has strong impact on the PE peak timing. According to our expectations, we observe an advancement or retardation of the PE by approximately 5~ps for application of a control pre- or post-pulse, respectively. We emphasize that the temporal shift is independent of the exact moment of control pulse application within the time range 1 (pre-pulse in Fig.~\ref{fig2}(d)) or 2 (post-pulse in Fig.~\ref{fig2}(e \& f)). In addition, we note that the duration of the detected PE signal remains constant with a FWHM of about 6.5~ps. Its amplitude is reduced by about 50\% which is attributed mainly to the damping of Rabi oscillations as will be discussed at the end of this section.

Due to the strong inhomogeneous broadening of the optical transitions, the PE temporal profile depends sensitively on the intensity of the \nth{1} excitation pulse. In particular, previous studies on (In,Ga)As QDs demonstrated that the PE signal acquires a non-Gaussian shape and experiences a significant advancement for pulse areas larger than $\pi$ due to inhomogeneity-induced dephasing of the excitons during the optical excitation~\cite{RabiQD16}. Therefore, the area of the \nth{1} pulse should be adjusted accurately to not too high values.

Figure~\ref{fig3} shows two-dimensional plots of the PE transients (horizontal axis) versus pulse area $\Theta_1 \propto \sqrt{I_1}$ (vertical axis). The top and bottom row show experimental data and numerical simulations, respectively. The simulations show $|P_{\mathrm{Signal}}|$, obtained from numerical solution of the extended OBE \cite{RabiQW17,Allen-Eberly}. These equations take into account the inhomogeneous broadening and the finite spot size of the laser pulses (for details, see Methods) and allow us to reproduce the measured coherent transients well. The results are presented for $ \tau_{12} = 80 $ ps and $\Theta_2 \approx \pi$ in the presence of a pre-pulse (Fig.~\ref{fig3}(a)) or a post-pulse (c) with area $\Theta_C \approx 2\pi$ and delays relative to the \nth{1} or \nth{2} pulse of 33~ps and 27~ps, respectively. The reference data without control pulse is shown in Fig.~\ref{fig3}(b). Cleraly, oscillations of the PE amplitude are observed with increasing $\Theta_1$ that have to be attributed to Rabi oscillations ~\cite{Malinovsky}.

In the simplest case of a TLS ensemble excited by a sequence of $\delta$-pulses, the PE peak is centered at $t=2\tau_{12}$ and its amplitude scales like $P_{\rm PE}\propto \sin(\Theta_1)$. In our experimental setting, the FWM signal at $t=2\tau_{12}$ is proportional to $|P_{\rm PE}|$. It follows from Fig.~\ref{fig3}(b) that we observe up to two full Rabi flops ($2\pi$ rotations). In our measurement, each maximum as function of $\Theta_1$ is attributed to $\Theta_1=m\pi/2$, where $m$ is an odd integer. In addition to the oscillatory behavior, we observe significant changes in the timing of the PE when $\Theta_1$ is varied. This is well reproduced by our numerical simulation (see Fig.~\ref{fig3}(e)) and indicates that dephasing of the ensemble during excitation with the \nth{1} and \nth{2} pulses should be taken into account~\cite{RabiQD16}.

The main result of Fig.~\ref{fig3} becomes obvious when comparing the plots with and without control pulse. Control pulse application shifts the PE signals along the time axis. Even though the PE transient profile may have a complex temporal shape, the impact of pre- or post-pulse with area $2\pi$ advances or retards the intensity-dependent transient, without strong changes in its shape. Thus, PE timing by the control pulse is demonstrated as confirmed for arbitrary intensities of the \nth{1} pulse. Moreover, in our experiment where excitation and control pulses have the same duration the inhibition of dephasing for high intensity pulses may add up. For example, it follows from Fig.~\ref{fig3}(a) that for $\Theta_1 \approx 3\pi/2$ and $\Theta_C \approx 2\pi$, the PE appears 8~ps earlier compared to the PE observed for $\Theta_1 \approx \pi/2$ and in absence of this control pulse. This advancement noticeably exceeds the PE pulse duration of 4~ps. 

\begin{figure}[h]
	\includegraphics[width=\columnwidth]{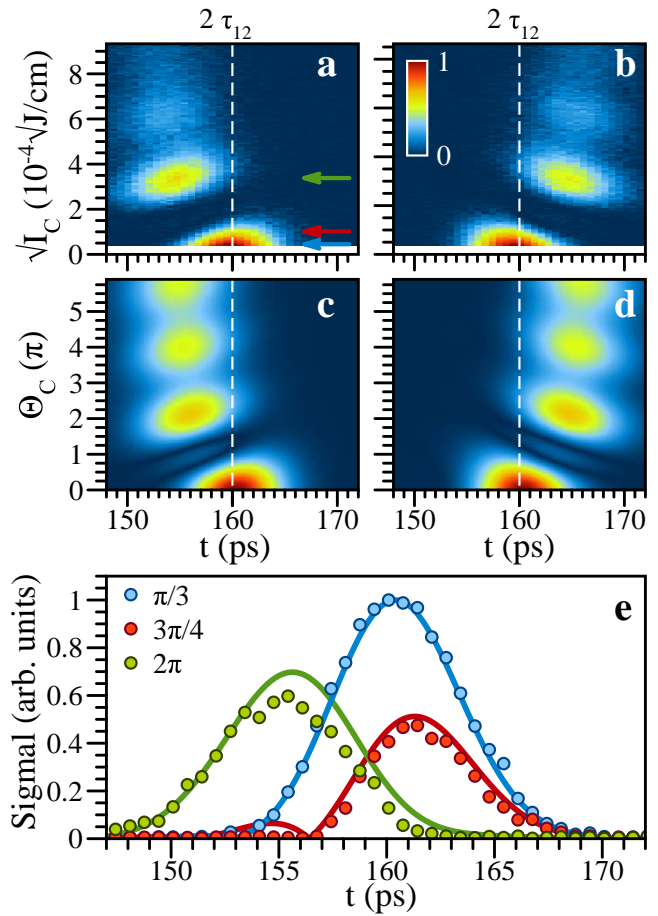}
	\caption{{\bf Dependence on control pulse area} Temporal PE profile as function of the control pulse area $\propto \sqrt{I_C}$: (a) \& (b) experimental data for contol pre- and post-pulses, (c) \& (d) corresponding simulations. The delay times for the pre- or post-pulses correspond to 33 and 27~ps, respectively, $\tau_{12} = 80$~ps, $\Theta_1 \approx \pi/2$, $\Theta_2 \approx \pi$. (e) Temporal PE profiles for three different pre-pulse intensities $I_C$ corresponding to $\Theta_C = \pi/3$ (blue), $3\pi/4$ (red) and $2\pi$ (green). Symbols and lines correspond to experiment and modeling, respectively. The associated intensities are labeled with arrows in (a).}
	\label{fig4}
\end{figure}

Finally, we discuss the influence of the area of the control pulse $\Theta_C$. Figure~\ref{fig4} shows the dependence of the PE temporal profile on the control pulse amplitude $\sqrt{I_C}$ for control  pre-pulse (a) and post-pulse (b) application, whereas (c) and (d) show the corresponding simulations, where $|P_{\mathrm{Signal}}|$ was calculated (see Methods for details). The time delays for pre- or post-pulses correspond to 33~ps and 27~ps relative to the \nth{1} or \nth{2} pulse, respectively. Here, we also observe oscillatory behavior due to Rabi flopping. PE timing shifts occur in both cases but the induced delays are opposite for pre- or post-pulse, so that the contour plots are mirrored with respect to $t=2\tau_{12}$. We concentrate on the PE timing in case of a pre-pulse. For weak pulse energies $\sqrt{I_C} < 1.5\times10^{-4}{\rm J^{1/2}/cm}$  ($\Theta_C < \pi $) we observe quenching of the PE amplitude, while its maximum slightly shifts towards longer times. For a pre-pulse amplitude of $\sqrt{I_C} \approx 1.5\times10^{-4}~{\rm J^{1/2}/cm}$ ($\Theta_C \approx \pi $) the PE almost disappears. Further increase of $I_C$ leads to the appearance of a strong PE signal with its maximum at $\sqrt{I_C} \approx 3\times10^{-4}~{\rm J^{1/2}/cm}$ ($\Theta_C \approx 2 \pi $), which is temporally advanced by 5~ps. The next maximum around $\sqrt{I_C} = 6 \times 10^{-4}~{\rm J^{1/2}/cm}$  ($\Theta_C \approx  4 \pi $)  is sufficiently weaker compared to the first one, and is only slightly further advanced. Thus, we conclude that the application of a control pre-pulse shows two different regimes depending on its pulse/intensity area. For $\Theta_C < \pi$ the maxima shift slightly to longer times, while application of a more intense control pulse $\Theta_C > \pi$ leads to an advanced PE and, in fact, a much larger temporal displacement (see Fig.~\ref{fig4}e).

The results of the simulations are in excellent agreement with the experimental data in Figs.~\ref{fig3} \& \ref{fig4}, considering the temporal shifts of the PE as well as their magnitude for pulse areas $\Theta_1$ \& $\Theta_C$ up to $2\pi$. The model accounts for the spatially inhomogeneous distribution of the laser intensity on the sample which leads to a spread of Rabi frequencies within the  laser spot which consequently leads to a decay of the Rabi oscillations~\cite{RabiQW17}. For pulse areas larger than $2\pi$ the damping of the Rabi oscillations is not reproduced quantitatively by our simulations, predominantly due to the simplified description of the laser intensity distributions. It is established that an intensity-dependent damping of the Rabi oscillations can also arise from the coupling to acoustic phonons as demonstrated for spatially homogeneous systems \cite{Kruegel-APB,Monniello-PRL}. However, in our case, the spatial inhomogeneity is likely the dominant damping process. A more detailed discussion of intensity-dependent damping is provided in the appendix section~\ref{Sec:DampingMech}.

Figure~\ref{fig4} shows that substantial PE shifts in time occur for control pulse areas up to $\Theta_C=2\pi$. For larger areas the PE shifts saturate, in agreement with our simulations which confirm that there exists a maximum possible shift $\tau_{\rm PE}$ for a given pulse duration $t_C$. However, $\tau_{\rm PE}$ is linearly proportional to $t_C$. The exact dependence of the PEs on intensity and duration of the control pulse is determined by the shape of its envelope. For long control pulses the PE intensity decreases, while the PE duration increases, probably because the freezing mechanism is not equally effective for the entire ensemble.

It should be noted that also intense off-resonant control pulses can be used to influence the dephasing of excitons in QD ensembles. In this case the ac Stark effect may lead to modulations of the PE amplitude as recently demonstrated in rare earth solids~\cite{Chaneliere2015,Ham2017,Faraon2018}. This behavior has also been confirmed within our model where we find, that the temporal shift of the PE decreases in presence of a detuning, for which also the PE amplitude rapidly reduces with increasing intensity of the control pulse.

Thus, we conclude that the timing of PEs by optical freezing of exciton dephasing and rephasing in QDs is robust with respect to the exact time of arrival of the control pulse. The PE shifts for a large variety of excitation conditions, e.g., it occurs for arbitrary intensities of the \nth{1} and the \nth{2} pulses and it is also independent of the optical phase of the control pulse. Therefore,  optical freezing can be applied for optical control of PE timing on ultrafast time scales, e.g. for timing corrections in memory protocols. Furthermore, addressing the spin degrees of freedom in systems with more than two energy levels ($V$- and $\Lambda$-type arrangements) can be used to split the temporal profile of PEs in two counter-polarized pulses making it possible to perform wave-function interferometry of electronic excitations in solid state systems. This opens a vast novel possibilities for investigations of collective phenomena in exciton ensembles.

\section{Acknowledgements}

Support by Dr. S. Maier in the epitaxial growth of the sample is gratefully acknowledged. We are grateful to D.~Suter and C.~Schmidt for useful discussions. We acknowledge financial support from the Deutsche Forschungsgemeinschaft (DFG) through the Collaborative Research Centre TRR 142 (project number 231447078 , project A02). S.V.P. thanks the Russian Foundation for Basic Research (Project No. 19-52-12046) and the Saint Petersburg State University (Grant No. 51125686). The W\"urzburg group acknowledges financial support by the state of Bavaria.

\section{Methods}

The experiment was performed on a single layer of (In,Ga)As QDs, surrounded by an AlGaAs $\lambda$-microcavity. The QD density is $1.8 \times 10^9 {\rm cm}^{-2}$. There is a Si $\delta$-layer with donor density $8 \times 10^9 {\rm cm}^{-2} $ located in the barrier 10~nm below the QD layer. The Bragg mirrors consist of alternating  GaAs and AlAs layers with thicknesses of 68~nm and 82~nm, respectively. The top and bottom mirror contain 5 and 18 of such pairs. The same sample was previously studied in Ref.~\cite{RabiQD16}. The sample has a gradient axis along which the thickness of the cavity and of the layers in the Bragg structure slightly change. Thereby, the energy of the photonic resonator mode varies in the spectral range of  $1.343 - 1.362 $~eV ($910-923$~nm) with a gradient of about 3.5 nm/mm. This spectral range corresponds approximately to the energy range of the photoluminescence from the QD ensemble. Therefore, the investigated structure shows strong photoluminescence with the maximum centered at the photon energy of the photonic mode and a full-width at half maximum (FWHM) of 7.4~meV,  resulting in the quality factor of about 200 (see Fig.~\ref{fig2}(b)).

The sample was kept in a bath cryostat and cooled down to 2~K. As source of the excitation and control pulses we use a single mode-locked Ti:sapphire laser, which generates pulses with tunable central wavelength at a repetition rate of 75.75~MHz and duration of 2.5~ps, estimated from autocorrelation measurements (see appendix section~\ref{Sec:PE-duration}). The spectral width of the laser pulses (FWHM = 0.6~meV) is below the spectral width of the cavity mode and therefore the resonator impact can be considered mostly as enhancement of the electric field amplitude~\cite{RabiQD16, Kasprzak}. Degenerate transient four-wave mixing (FWM) was measured in reflection geometry (see Fig.~\ref{fig2}(a)). The \nth{1} and \nth{2} pulses hit the sample with wave vectors $\textbf{k}_1 $ and $\textbf{k}_2 $ at incidence angles of $3 ^{\circ}$ and $4 ^{\circ}$, respectively, and with the delay time $ \tau_{12}$ between them. In addition, we applied a control pulse, which hits the sample with the wave vector $\textbf{k}_2 $, delayed relative to the \nth{1} or \nth{2} pulse. The delay times between the pulses are controlled by mechanical translation stages and their polarizations are set by Glan prisms combined with half-wave-plates.

The laser beams are focused at the sample into spots with diameters of about 400~$\mu$m for the \nth{1} pulse and about 250~$\mu$m for the \nth{2} and control pulses. 
In order to record PEs from the charged QDs only, the \nth{1} pulse had a horizontal (H) polarization, while the \nth{2} and control pulses were vertically (V) polarized. The H-component of the signal was detected along the $ 2\textbf{k}_2 -\textbf{k}_1$ direction. The FWM signal was overlapped with the reference pulse and the resulting interference signal was heterodyne-detected at the balanced photoreceiver~\cite{FTTrev18}. The delay line of the reference pulse $t_{\rm ref}$ was scanned relative to the \nth{1} pulse to obtain the temporal profile of the electric field amplitude of the FWM signal. As result, the cross correlation of the electrical field amplitude was detected (see Eq.~(\ref{eq:convolution}) in the next section). The area of the \nth{2} pulse was fixed at $\pi$, while the areas of the exciting and control pulses were varied.

\textbf{Theoretical modeling.} The photoexcited TLS are theoretically described by the optical Bloch equations (OBE) \cite{Allen-Eberly,Meier} Their solution provides the dynamics of the microscopic polarization $p=\langle 2|\hat{\rho}|1\rangle$ and the occupation of the upper energy level $n=\langle 2|\hat{\rho}|2\rangle$ where $\hat{\rho}$ is the density matrix of the system~\cite{Meier}. Taking into account the inhomogeneous broadening of the TLS ensemble and the finite spot size of the laser pulses leads to a set of extended OBE \cite{RabiQW17}
\begin{align}
\frac{\partial}{\partial t} p_{i}(\boldsymbol{r},t)=& -(1/T_2+i\omega_i) p_{i}(\boldsymbol{r},t)\notag\\
&+(i/\hbar)d_{12} \E(\boldsymbol{r},t)(1-2n_{i}(\boldsymbol{r},t)),\\
\frac{\partial}{\partial t} n_{i}(\boldsymbol{r},t)=&-n_{i}(\boldsymbol{r},t)/T_1-2(d_{12}/\hbar) \mathrm{Im}[p^*_{i}(\boldsymbol{r},t)\E(\boldsymbol{r},t))],
\label{eq:OBE}
\end{align}
where $T_2$ and $T_1$ are the coherence time and lifetime of the exciton, respectively. The index $i$ describes the inhomogeneous broadening of the resonance as a superposition of many TLS with different transition frequencies $\omega_i = (E_{2,i} - E_{1,i})/\hbar$. The fraction of TLS with transition frequency $\omega_{i}$ is described by the weight function $G(\omega_{i})$, which is taken to be a Gaussian with FWHM $\hbar\Delta_0 = 7.5~\mathrm{meV}$. $\E(\boldsymbol{r},t) = \E_1(\boldsymbol{r},t) + \E_2(\boldsymbol{r},t) + \E_c(\boldsymbol{r},t)$ is the total electric field amplitude  including the \nth{1}, \nth{2}, and control pulses. The temporal dependences of the pulse amplitudes are modeled as Gaussians with FWHM of $\Delta t = 2.5~\mathrm{ps}$, while their magnitudes are determined by the pulse areas, Eq.~(\ref{pulse-area}). The finite spot size of the laser pulses is included by considering their spatial profiles as function of $\boldsymbol{r}=(x,y)$. These profiles are taken as
\begin{align}
\E(\boldsymbol{r},t) = \exp(-r^2/\sigma^2_R)(\E_1(t) + \E_2(t) + \E_c(t)),
\end{align}
where $r=\sqrt{x^2+y^2}$ is the distance from the laser spot center at $(0,0)$, and $\sigma_R$ is the spot width. For simplicity, we assume that all spots have the same center and the same diameter. To obtain the macroscopic polarization $P(\boldsymbol{r},t)$ that is created in the sample, the sum of all microscopic polarizations $p_i(\boldsymbol{r},t)$, weighted with $G(\omega_i)$ and multiplied with $d_{12}$ has to be calculated according to
\begin{align}
P(\boldsymbol{r},t) = d_{12} \sum_{i} G(\omega_i) p_i(\boldsymbol{r},t).
\end{align}

The macroscopic polarization $P(\boldsymbol{r},t)$ is the source of the electric field that is emitted from the spot at position $\boldsymbol{r}$. The calculations can be performed with $P(\boldsymbol{r},t)$, since its absolute value is proportional to the absolute value of the emitted electric field. The detector does not resolve each coordinate separately, but rather, the spatial integral of all signals is measured, which requires spatial averaging. However, before the electric field that is induced by the macroscopic polarization $P(\boldsymbol{r},t)$ is detected, it interferes with the reference pulse $\E_{\mathrm{Ref}}(\boldsymbol{r},t)$, this interference is described by a temporal convolution of the macroscopic polarization and the electric field of the reference pulse that needs to be performed for each coordinate separately. The reference pulse is assumed to have the same spatial profile as $\E(\boldsymbol{r},t)$, which results in a factor of $\exp(-r^2/\sigma^2_R)$ for each $P(\boldsymbol{r},t)$. We note that the convolution of two functions is a linear operation, which means that we can perform the spatial averaging first and calculate the temporal convolution in the end. The total signal that results from the spatial averaging then reads
\begin{align}
P_{\mathrm{average}}(t) = \int \mathrm{d}\boldsymbol{r} \exp(-r/\sigma^2_R) P(\boldsymbol{r},t).
\label{eq:totalFWM}
\end{align}

Due to the rotational symmetry of all spatial profiles, it is advantageous to use a polar coordinate system. With this, the extended OBE only need to be solved for all distances $r$ instead of all coordinates. Discretizing the distance $r$ with the index $s$, and therefore changing the integration to a summation, leads to the following expression
\begin{align}
P_{\mathrm{average}}(t) = 2\pi \sum_{s} r_s \Delta r \exp(-r_s/\sigma^2_R) P(r_s,t),
\label{eq:totalFWM}
\end{align}
where $\Delta r$ is the stepwidth for the discretization of $r$.


The final signal $P_{\mathrm{Signal}}(t)$ is taken by calculating the temporal convolution of $P_{\mathrm{average}}(t)$ and the temporal dependence of the reference pulse $\E_{\mathrm{Ref}}(t)$
\begin{align}
P_{\mathrm{Signal}}(t) = (P_{\mathrm{average}} * \E_{\mathrm{Ref}})(t),
\label{eq:convolution}
\end{align}
where $\E_{\mathrm{Ref}}(t) = \E_1(t) / \mathrm{max\{\E_1(t)\}}$ is given by the normalized first pulse
and the convolution of two functions $f$ and $g$ is defined as
\begin{align}
(f * g)(x) := \int_{-\infty}^{\infty} f(y)g(x-y)\mathrm{d}y.
\label{eq:convolutiondef}
\end{align}
For the implementation of the inhomogeneous ensemble, $1500$ TLS are considered with frequencies from $-15~\mathrm{meV}$ to $15~\mathrm{meV}$ and a resolution of $0.02~\mathrm{meV}$. For the spatial profile of the laser pulses, the value of $\sigma_R$ is not needed in the calculation. Rather, the radii can be written in units of $\sigma_R$. Radii from $r=0.05\sigma_R$ to $r=3.5\sigma_R$ are considered with a stepwidth of $\Delta r = 0.05\sigma_R$. The coherence time and lifetime of the exciton are chosen according to the experimental values of $T_2=710~\mathrm{ps}$ and $T_1=360~\mathrm{ps}$. The extended OBE are integrated as function of time with the fourth-order Runge-Kutta method, using a temporal stepwidth of $0.01 ~\mathrm{ps}$. For scanning the pulse area, a stepwidth of $\Delta\Theta=0.05$ was used.

\appendix

\section{Freezing of dephasing for rectangular control pulse}
\label{Sec:Rect}

Let us consider an ensemble of two-level systems (TLS) resonantly excited with rectangular pulses. During the excitation process, the nutation of the Bloch vector $\mathbf{U}$ around the vector $\mathbf{\Omega} = (\Omega_R,0,\Delta)$ is given by ~\cite{Allen-Eberly, Malinovsky}

\begin{widetext}
\begin{equation}
\label{Eq:nutation}
\left(\begin{array}{c} u(t) \\ v(t) \\ w(t) \end{array}\right) 
=
\left(
\begin{array}{ccc} 
         \frac{\Omega_R^2+\Delta^2\cos(\widetilde{\Omega} t)}{\widetilde\Omega^2} & -\frac{\Delta}{\widetilde{\Omega}}\sin{\widetilde\Omega t} & ~~\frac{\Omega_R \Delta}{\widetilde\Omega^2}\left( 1- \cos{\widetilde\Omega t} \right) \\ 
         \frac{\Delta}{\widetilde\Omega}\sin(\widetilde\Omega t) & 
         \cos{\widetilde\Omega t} & \frac{\Omega_R}{\widetilde\Omega}\sin{\widetilde\Omega t} \\
         \frac{\Omega_R \Delta}{\widetilde\Omega^2}\left( 1- \cos{\widetilde\Omega t} \right) & ~~\frac{\Omega_R}{\widetilde\Omega}\sin(\widetilde\Omega t) & \frac{\Omega_R^2\cos(\widetilde{\Omega} t)+\Delta^2}{\widetilde\Omega^2} 
         \end{array} \right)
\left(\begin{array}{c} u_0 \\ v_0 \\ w_0 \end{array}\right) ,
\end{equation}
where $\Delta$ is the detuning frequency, $\Omega_R$ is the Rabi frequency and  $\widetilde\Omega = \sqrt{\Omega_R^2+\Delta^2}$ is the generalized Rabi frequency.  Before the action of the control pulse all Bloch vectors are distributed along the equator ($w_0=0$). Their initial coordinates can be expressed through the phase $\phi$, i.e. $u_0=\sin{\phi}$ and $v_0=\cos{\phi}$.  Setting $\Omega_R t_C = 2 \pi n$ ($n$ is an integer) and taking into account that $\Omega_R \gg \Delta$ we obtain the coordinates of a particular Bloch vector at the end of the pulse action $t=t_C$ 
\begin{equation}
\label{Eq:coordinates}
\left(\begin{array}{c} u(t_C) \\ v(t_C) \\ w(t_C) \end{array}\right) 
=  \left(\begin{array}{c} \sin{\phi} - \frac{\Delta}{\Omega_R}\sin{\left[\frac{\Delta}{2\Omega_R}  t_C \Delta\right]}\cos{\phi} \\  
\frac{\Delta}{\Omega_R}\sin{\left[ \frac{\Delta}{2\Omega_R}t_C \Delta\right]}\sin\phi + \cos{\left[ \frac{\Delta}{2\Omega_R}t_C \Delta\right]}\cos{\phi}\\ 
\sin{\left[ \frac{\Delta}{2\Omega_R}t_C \Delta\right]}\cos{\phi}  \end{array}\right).
\end{equation}
\end{widetext}
Here, we neglected the terms with the order higher than $(\Delta/\Omega_R)^2$.

We consider the deviation $\mathbf{\Delta U} = \mathbf{U(t_C)}-\mathbf{U_0}$ where 
$\mathbf{U_0}=(\sin{\phi},\cos{\phi},0)$. If we assume that $\frac{\Delta}{2\Omega_R}t_C \Delta \ll 1$, we obtain the following expression for the deviation of each coordinate
\begin{eqnarray}
  \delta u \approx -\frac{\cos{\phi}}{2}\left(\frac{\Delta}{\Omega_R}\right)^2 ~t_C \Delta, \\
  \delta v \approx \left[ \frac{\sin{\phi}}{2}-\frac{\cos{\phi}}{8}t_C \Delta \right] \left(\frac{\Delta}{\Omega_R}\right)^2 t_C \Delta,  \\
  \delta w \approx \frac{\cos{\phi}}{2} \frac{\Delta}{\Omega_R} t_C \Delta.
\end{eqnarray}
It follows that the deviation of the phase after action of the pulse is $\delta\phi \approx \sqrt{\delta u^2 + \delta v^2} < \left(\Delta/\Omega_R \right)^2 t_C \Delta$ for any phase $\phi$ and $t_C\Delta<4$. The deviation of the colatitude $\delta\theta \approx \delta w <  \frac{1}{2} \left(\Delta/\Omega_R\right) t_C \Delta$ is larger as compared to $\delta\phi$. Nevertheless, all changes can be considered as small compared to $t_C\Delta$. This is valid also when the duration of the control pulse is large, i.e.  $t_C\Delta > 4$ as long as $\frac{\Delta}{2\Omega_R}t_C \Delta \ll 1$ holds.

\section{Evaluation of PE pulse duration}
\label{Sec:PE-duration}

\begin{figure*}[ht]
	\includegraphics[width=1.5\columnwidth]{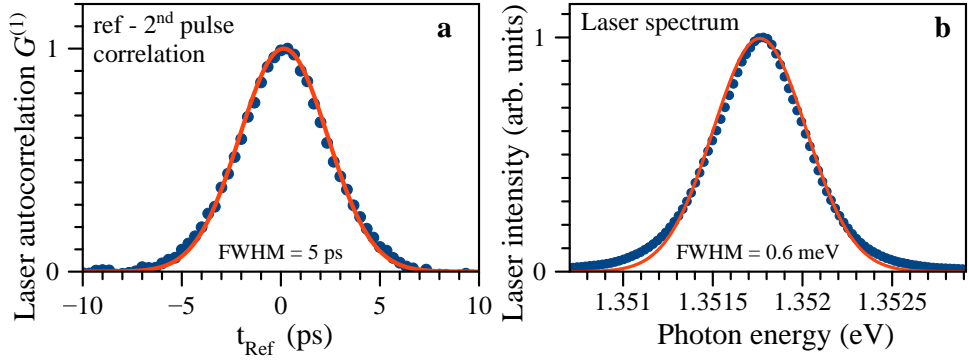}
	\caption{{\bf Properties of the laser pulse}: (a) envelope profile of the electric field auto-correlation function, (b) intensity spectrum.}
	\label{Fig1}
\end{figure*}

Figure~\ref{Fig1}(a) shows the envelope of the electric field auto-correlation function for the laser pulse used in the experiment. The auto-correlation measurements were performed using the heterodyne signal resulting from interference of the \nth{2} $E_2(t)$ and reference $E_{\rm Ref}(t)$ pulses:  $G^{(1)}(t_{\rm Ref}) \propto \int E_2^*(t) E_{Ref}(t-t_{\rm Ref}) dt$. The correlation is well described by a Gaussian with FWHM of 5~ps. Taking into account that all laser pulses are identical in our experiment, the electric field envelope of each pulse has a FWHM of 3.5 ps, which corresponds to the pulse duration of $\tau_d = 2.5$~ps. The intensity spectrum of the laser which is shown in Fig.~\ref{Fig1}(b) is in agreement with this evaluation. It can be approximated as well with a Gauss function with 0.6 meV FWHM, which corresponds to the pulse duration of approximately 3~ps.

Setting $\tau_d = 2.5$~ps we can estimate the PE pulse duration $\tau_{d,{\rm PE}}$ from the measured FWM signal, which is given by the cross-correlation of the FWM optical field $E_{\rm FWM}(t)$ with the reference field (see also Eq.~(9) in the Methods section). The signal is well described by a Gaussian with the FWHM $\tau_{cc,{\rm PE}}=$6.5~ps (see Fig.~2-4 in the main text). Assuming that the PE pulse is also Gaussian, we obtain $\tau_{d,{\rm PE}} = \sqrt{\tau_{cc,{\rm PE}}^2/2-\tau_d^2} \approx $~3.8~ps.

%
\section{Intensity-dependent damping mechanisms}
\label{Sec:DampingMech}
\begin{figure*}[ht]
	\includegraphics[width=1.5\columnwidth]{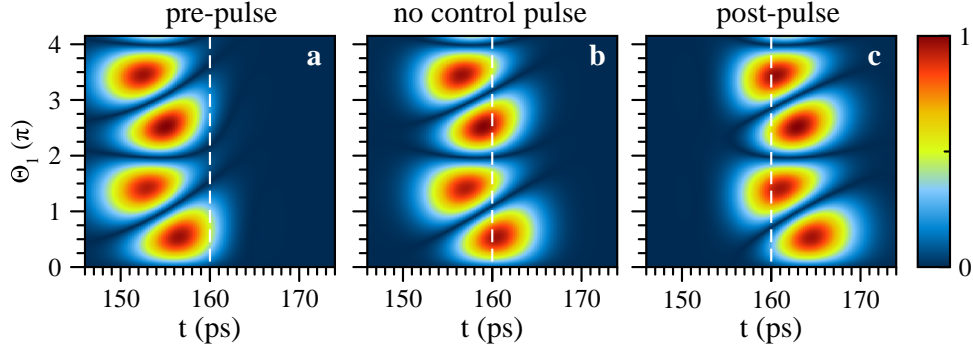}
	\caption{Calculated PE signals as function of time and $\Theta_1$ when the finite size of the laser spots is not taken into account. The simulations were performed in the presence of (a) a control pre-pulse, (b) no control pulse, and (c) a control post-pulse.}
	\label{damp-none}
\end{figure*}

As shown in the main text, the experimental data are well described by an inhomogeneously broadened ensemble of two-level systems when the spatial profiles of the laser pulses are taken into account. The intensity-dependent reduction of the PE amplitude in Fig.~3 and Fig.~4 is a consequence of the finite spot sizes of the laser pulses. If one instead assumes a spatially uniform excitation, see Fig.~\ref{damp-none}, the PEs remain undamped.

In section~\ref{sec-foci} we analyze these geometrical properties in more detail. In section~\ref{sec-phonons} we investigate another well established damping mechanism, namely the coupling to acoustic phonons, which for our situation turns out to be of minor importance.

%
%
\subsection{Different foci of the spots}
\label{sec-foci}
So far, we concluded that the spatial averaging of the spots of the incident laser pulses is important for describing the experimental data. However, the assumption that all pulses are focused on the same spot is an idealized scenario, which is difficult to realize experimentally and has only been used for the sake of simplicity. In the following we investigate what happens if the laser pulses do not have exactly the same spatial distributions of intensity. In particular, we consider one example when the spot sizes are identical while the focus of the first pulse is slightly displaced by an amount $x_0$ from the foci of the other pulses. In this case, we have no rotational symmetry anymore and the simple expression Eq.~(5) does no longer hold. Thus, the spot has to be fully spatially resolved and a cartesian coordinate system has to be used. The total electric field depending on $x$ and $y$ reads
\begin{align}
\E(\boldsymbol{r},t) = e^{-((x-x_0)^2+y^2)/\sigma^2_R} \E_1(t) + \nonumber\\
 e^{-(x^2+y^2)/\sigma^2_R}(\E_2(t) + \E_3(t)).
\end{align}
In this case, the total signal that results from the spatial averaging needs to be integrated along two dimensions. Discretizing $x$ and $y$ with the indices $s$ and $s'$, respectively, leads to the following expression
\begin{align}
P_{\mathrm{average}}(t) = \sum_{s,s'}  \Delta x \Delta y \exp(-(x_s^2+y_{s'}^2)/\sigma^2_R) P((x_s,y_{s'}),t),
\label{eq:totalFWM}
\end{align}
from which we obtain $P_{\mathrm{Signal}}(t)$ by performing the temporal convolution with the reference pulse. For the simulation, the spot was sampled from $x=-3.5\sigma_R$ to $x=5\sigma_R$ and from $y=-3.5\sigma_R$ to $y=3.5\sigma_R$ with a stepwidth of $\Delta x = \Delta y = 0.2\sigma_R$. While all parameters were kept the same, the stepwidths were increased due to the large amount of calculated points. The temporal stepwidth is chosen to be $0.04 ~\mathrm{ps}$ and the stepwidth of the pulse areas to be $\Delta\Theta=0.3$. The displacement is chosen to be $x_0=0.5\sigma_R$.

\begin{figure*}[ht]
	\includegraphics[width=1.5\columnwidth]{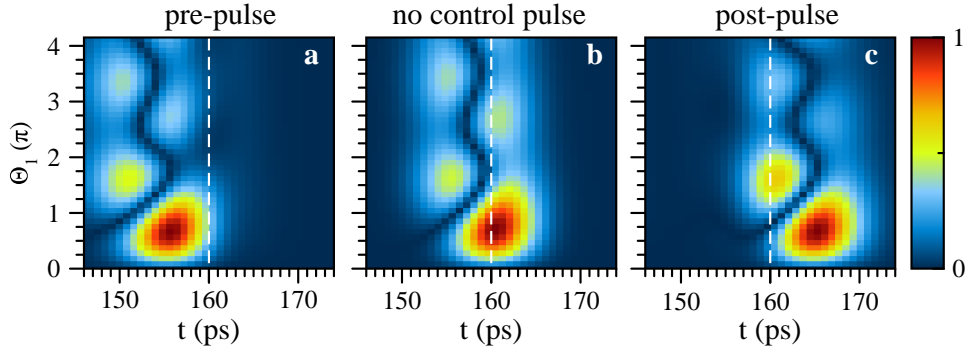}
	\caption{Calculated PE signals as function of time and $\Theta_1$. Here, the finite size of the laser spots and a displacement of the first pulse by $x_0=0.5\sigma_R$ was considered. The simulations were performed in the presence of (a) a control pre-pulse, (b) no control pulse, and (c) a control post-pulse.}
	\label{damp-foci}
\end{figure*}
\begin{figure*}[ht]
	\includegraphics[width=1.5\columnwidth]{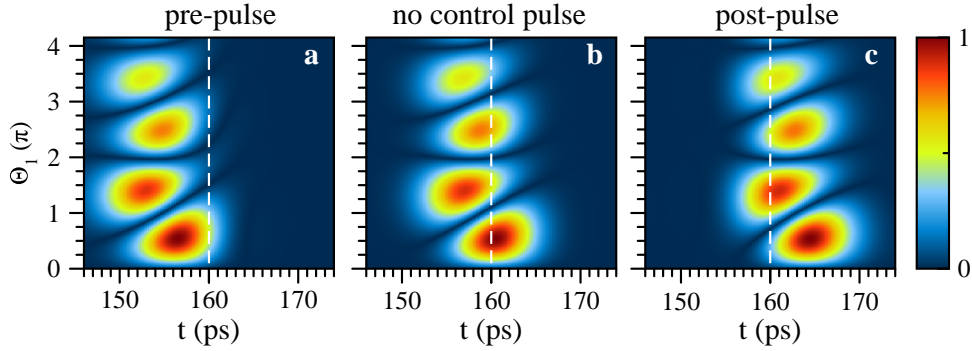}
\caption{Calculated PE signals as function of time and $\Theta_1$ considering the finite size of the laser spots and including an additional phonon-induced intensity-dependent dephasing. The simulations were performed in the presence of (a) a control pre-pulse, (b) no control pulse, and (c) a control post-pulse.}
	\label{damp-phon}
\end{figure*}

Fig.~\ref{damp-foci} shows the temporal evolution of the PE as function of the area of the first pulse for the control pre- and post-pulse scenarios, as well as for the scenario without control pulse. Comparing the plots with Fig.~3 in the main text, we find that non-identical foci of the laser spots may result in an additional intensity-dependent damping mechanism.
For the considered parameters, which are close to the experimental conditions, the intensity-dependent damping of the Rabi oscillations is similar to the measured data. Note that different foci of the  laser spots have no significant influence on the timing of the PE, i.e. the advancement and retardation of the PE pulse in the presence of a control pre- or post-pulse.


\subsection{Coupling to acoustic phonons}
\label{sec-phonons}

In literature, the damping of Rabi oscillations due to coupling to reservoirs - especially acoustic phonons - has been investigated thoroughly~\cite{phon-Mogilevtsev,Monniello-PRL,phon-Ramsay,Kruegel-APB}. A proper microscopic treatment of the interaction with phonons like, e.g., in Ref.~\cite{Kruegel-APB}, is beyond the scope of our  model and is also not directly applicable since we have to take into account the spatially-inhomogeneous optical excitation. Following Refs.~\cite{Monniello-PRL,phon-Ramsay} we rather assume that an additional intensity-dependent dephasing process occurs as long as the laser fields are switched on. Therefore, we change Eq.~(3) to 
\begin{align}
\frac{\partial}{\partial t} p_{i}(\boldsymbol{r},t)= -(1/T_2 + \kappa\Omega_R^2(\boldsymbol{r},t)+ i\omega_i) p_{i}(\boldsymbol{r},t)+ \nonumber \\
+(i/\hbar)d_{12} \E(\boldsymbol{r},t)(1-2n_{i}(\boldsymbol{r},t))
\end{align}
and incorporate the effective exciton-phonon interaction into the damping function $\kappa$.
In order to identify mere phonon-induced damping effects no spatial averaging is applied here. The results for this simplified model are shown in Fig.~\ref{damp-phon}. Due to the low temperature of 2~K in our experiment the proportionality coefficient in the damping function $\kappa$ is chosen to be four times larger than that used in Ref.~\cite{Monniello-PRL}, in order to obtain visible effects.

Indeed, the additional term leads to some damping of the PE signal. However, even though the coupling has been artificially increased by a factor of four, the intensity-dependent decrease is much weaker than in the experiment and when compared to the simulations that include the spatial profiles of the laser spots. Therefore, we are convinced that the geometrical averaging over the laser spots is the by far most relevant contribution to the damping at higher excitation intensities for the presented experimental data.

\end{document}